\documentclass[aps,pra,twocolumn,showpacs,superscriptaddress,groupedaddress]{revtex4}  

\usepackage{graphicx}   
\usepackage{dcolumn}   
\usepackage{bm}             
\usepackage{amssymb}   
\usepackage{epsfig}
\usepackage{color}

\begin{document}


\title{Dissociative electron attachment and electron-impact resonant dissociation of vibrationally excited O$_2$ molecules}

\author{V. Laporta}
\email{vincenzo.laporta@imip.cnr.it}
\affiliation{Istituto di Metodologie Inorganiche e dei Plasmi, CNR, Bari, Italy}
\affiliation{Department of Physics and Astronomy, University College London, London WC1E 6BT, UK}

\author{R. Celiberto}
\affiliation{Dipartimento di Ingegneria Civile, Ambientale, del Territorio,
Edile e di Chimica, Politecnico di Bari, Italy}
\affiliation{Istituto di Metodologie Inorganiche e dei Plasmi, CNR, Bari, Italy}

\author{J. Tennyson}
\affiliation{Department of Physics and Astronomy, University College London, London WC1E 6BT, UK}

\begin{abstract}
State-by-state cross sections for dissociative electron attachment and electron-impact dissociation for molecular oxygen are computed using \emph{ab initio} resonance curves calculated with the R-matrix method.  When O$_2$ is in its vibrational ground state, the main contribution for both processes comes from the $^2\Pi_u$ resonance state of O$^-_2$ but with a significant contribution from the $^4\Sigma^-_u$ resonant state. Vibrational excitation leads to an increased contribution from the low-lying $^2\Pi_g$ resonance, greatly increased cross sections for both processes and the threshold moving to lower energies. These results provide important input for models of O$_2$-containing plasmas in non-equilibrium conditions.
\end{abstract}

\maketitle

\section{\label{In} Introduction}
Molecular oxygen is a major component of the Earth's atmosphere and is fundamental  for life. It plays an important role in many natural and technological processes. In this paper we focus on the ability of low-energy electrons to dissociate molecular oxygen, in its fundamental electronic state, by means of resonant scattering. In particular we consider the state-by-state processes of dissociative electron attachment (DEA),
\begin{equation}
e^- + \textrm{O}_2(\textrm{X}\,^3\Sigma^-_g; v) \to \textrm{O}_2^{-*}
\to \textrm{O}^-(^2\textrm{P}) + \textrm{O}(^3\textrm{P})\,, \label{eq:DeA_process}
\end{equation}
and electron impact dissociation (EID),
\begin{equation}
e^- + \textrm{O}_2(\textrm{X}\,^3\Sigma^-_g; v) \to \textrm{O}_2^{-*} \to e^- +
2\,\textrm{O}(^3\textrm{P})\,, \label{eq:Diss_process}
\end{equation}
for each oxygen vibrational level $v$. The threshold for the process (\ref{eq:DeA_process}) is 3.64 eV, corresponding to the asymptotic energy of O and O$^-$ fragments from the level $v=0$, whereas for process (\ref{eq:Diss_process}) corresponds to the oxygen dissociation of 5.11 eV.

Such collisions are important to study phenomena occurring in the upper atmosphere, re-entry physics, electrical discharges and plasma chemistry. DEA is the principal mechanisms, in molecular plasmas, for forming negative ions from neutral molecules; the inverse process represents associative detachment. Although significant theoretical and experimental effort has been invested in characterizing electron-O$_2$ cross sections~\cite{itikawa:1}, information is only available for the vibrational ground state of O$_2$ for which both processes have rather small cross sections. Here, we will show that DEA and EID processes become much more important for vibrationally excited oxygen, as the corresponding cross sections increases by orders of magnitude with the excitation of the molecule. These cross sections represent fundamental input quantities in kinetics models of oxygen-containing non-equilibrium plasmas, where high-lying vibrational levels of molecules can be hugely populated.

The paper is organized as follow: In the Section~\ref{TM} we will give a brief account and numerical details on the adopted theoretical models for the calculation of the potential energy curves and on the description of the nuclear dynamics. The results are illustrated in Section~\ref{R_D} while the concluding remarks are given in Section~\ref{Con}.

\section{\label{TM} Theoretical model}
Many low-energy electron-molecule scattering processes are dominated by resonances. For molecular oxygen these processes can be quantitatively understood in term of four low-lying O$_2^-$ resonances of symmetry $^2\Pi_g$, $^2\Pi_u$, $^4\Sigma^-_u$ and $^2\Sigma^-_u$. First calculations were made by Noble \emph{et al.}~\cite{PhysRevLett.76.3534} who used the R-matrix method~\cite{Tennyson_PR_2010} to determine the positions and widths for these four resonances. The O$_2$ target was represented  using nine states, corresponding to the orbital configurations: $[core]1\pi^4_u1\pi_g^2$ and $[core]1\pi^3_u1\pi_g^3$.  The scattering $T$-matrix was calculated, in the fixed-nuclei approximation, for internuclear distance range $[1.85,\,3.5]\,a_0$, using the configurations: $[core]1\pi^4_u1\pi_g^3\,(^2\Pi_g)$, $[core]1\pi^3_u1\pi_g^4\,(^2\Pi_u)$, $[core]1\pi^4_u1\pi_g^23\sigma_u\,(^4\Sigma^-_u)$ and $[core]1\pi^4_u1\pi_g^23\sigma_u\,(^2\Sigma^-_u)$, for energies up to 15 eV.

In our previous paper on state-to-state cross sections for resonant vibrational excitation (RVE) for electron-oxygen 
scattering~\cite{0963-0252-22-2-025001}, we extended Noble \emph{et al.}'s calculations toward larger internuclear distances 
using the quantum chemistry code MOLPRO~\cite{MOLPRO_brief}. We used a multi-reference configuration interaction (MRCI) model and an 
aug-cc-pVQZ basis set. For the neutral ground-state of O$_2$ as well as for the real part of 
its anionic state O$_2^-$, the active space included the first 2 core orbitals (1$\sigma_g$, 2$\sigma_g$) and 
8 valence orbitals (2$\sigma_g$, 3$\sigma_g$, 2$\sigma_{u}$, 3$\sigma_{u}$, 1$\pi_{u}$,   1$\pi_{g}$)
plus 150 external orbitals for a total of 160 contractions. 
The MRCI orbitals were taken from a  ground-state multi-configuration self-consistent field (MCSCF) calculation in which 
4 electrons were kept frozen in the core orbitals and excitations among all the valence orbitals were considered for the 12- and 13-electrons 
of the neutral and ionic state respectively.

In the present work we update Noble \emph{et al.}'s R-matrix calculations in order to have a better representation of the states involved in the 
scattering processes. Calculations were performed using the UKRMol codes \cite{jt518}.
Orbitals for the O$_2$ were generated using configuration interaction (CI) calculations using a cc-pVTZ Gaussian-type orbital (GTO) basis set. 
The electrons in the lowest 3 core orbitals, (1$\sigma_g$, 2$\sigma_g$, 1$\sigma_{u}$)$^6$, were frozen  and an active space
was constructed from 
9 valence orbitals (3$\sigma_g$, 2$\sigma_{u}$, 3$\sigma_{u}$, 1$\pi_{u}$, 2$\pi_{u}$,  1$\pi_{g}$)$^{10}$. 
The scattering calculations with 17 electrons also used a GTO basis \cite{jt286} to represent the continuum electron. 
The calculations were based on use of a complete active space CI representation 
which involves placing the extra electron in both a continuum orbital and target orbitals using the following perscription:
(1$\sigma_g$, 2$\sigma_g$, 1$\sigma_{u}$)$^6$ (3$\sigma_g$, 2$\sigma_{u}$, 3$\sigma_{u}$, 1$\pi_{u}$, 2$\pi_{u}$,  1$\pi_{g}$)$^{11}$
 and (1$\sigma_g$, 2$\sigma_g$, 1$\sigma_{u}$)$^6$ (3$\sigma_g$, 2$\sigma_{u}$, 3$\sigma_{u}$, 1$\pi_{u}$, 2$\pi_{u}$,  1$\pi_{g}$)$^{10}$
(4$\sigma_g$, 5$\sigma_g$, 3$\pi_{u}$, 4$\sigma_u$, 2$\pi_{g}$, 3$\pi_{g}$, 1$\delta_g$, 1$\delta_u$)$^1$,
where the second set of configurations  involves placing the extra electron in an uncontracted \cite{jt189} target virtual orbital.
Calculations used an R-matrix radius of 10~a$_0$. Both the MOLPRO and R-matrix calculations were performed using D$_{2h}$ symmetry.

\begin{figure}
\includegraphics[scale=.8]{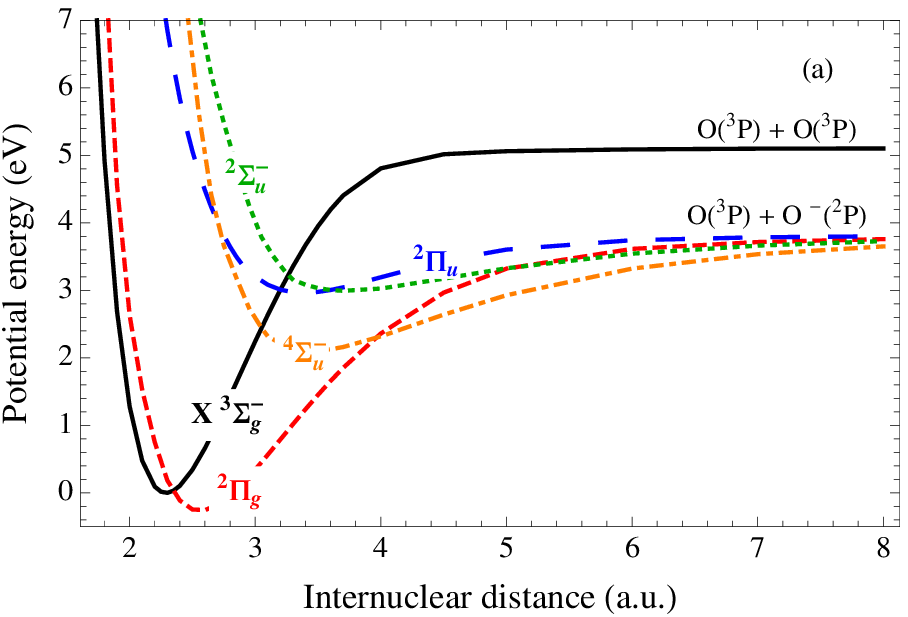}
\\
\includegraphics[scale=.8]{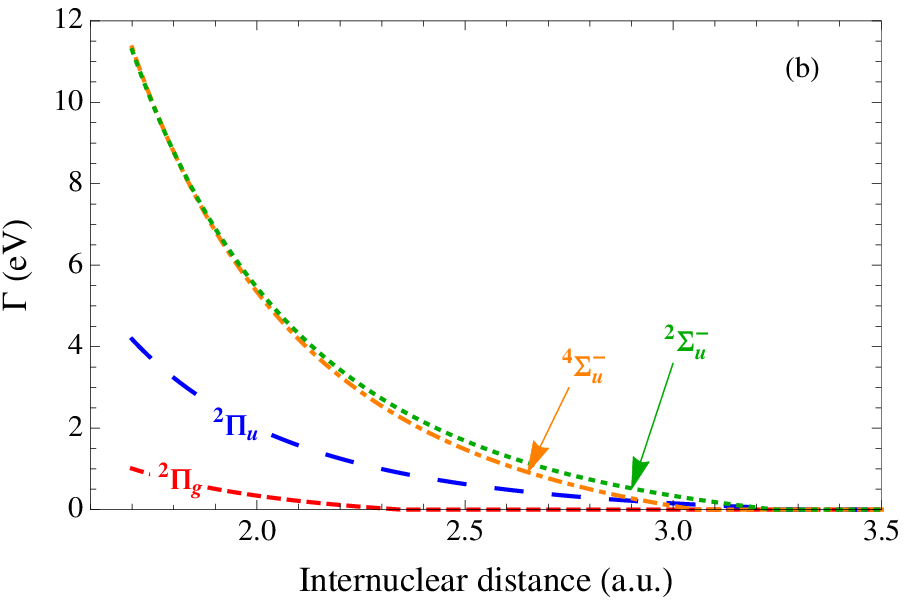}
\caption{(Color online) (a) Potential energy curves for ground state $\textrm{X}\,^3\Sigma_g^-$
of O$_2$ and for the four lowest resonant states O$_2^-$; (b) The corresponding
resonance widths, $\Gamma(R)$, as a function of the internuclear distance.
\label{fig:pes}}
\end{figure}

The potential curves calculated with the R-matrix and MOLPRO codes use different basis-sets, different theoretical models and computational methods. 
This means that the resulting curves did not join and a merging procedure was required. First of all the O$_2$ neutral ground-state potential
 was calculated using  MOLPRO and all the subsequent results were referred to this curve. The real part of the 
 resonance potentials, computed as a bound scattering state in the R-matrix calculations, were fixed 
at large internuclear distances beyond the crossing point to the MOLPRO curve which precisely  
reproduces the experimental oxygen atom electron affinity of 1.46 eV.
At shorter internuclear distances,  the R-matrix resonant curves, are simply a smooth
continuation of these large $R$ curves and therefore their positions are fixed by
the corresponding MOLPRO curve.
Potential energy curves for the O$_2(\textrm{X}\,^3\Sigma_g^-)$ ground state, which supports 42 vibrational states, and the four resonance states, 
plus the corresponding widths, are reported in Fig.~\ref{fig:pes}. We note that the new calculated potentials are slightly changed with respect 
to those of our previous paper~\cite{0963-0252-22-2-025001}, in particular for the $^2\Pi_u$ symmetry, and this could affect the RVE cross section 
values. To check this aspect, we performed new calculations for RVE processes, observing however no significant changes in the cross sections with respect to the corresponding results in~\cite{0963-0252-22-2-025001}.

Previously~\cite{0963-0252-22-2-025001} we showed that for energies up to about 4 eV the RVE cross sections 
are characterized by narrow spikes dominated by the $^2\Pi_g$ resonance. For energies about 10 eV, the cross sections 
showed a broad maximum due to enhancement by the $^4\Sigma^-_u$ resonance with a smaller contribution coming from the $^2\Sigma^-_u$ state. In the case of DEA and EID processes we show that they are dominated, at least for low vibrational levels $v$ of O$_2$, by the $^2\Pi_u$ resonance. Although electron attachment from oxygen molecules 
has been widely studied~\cite{McConkey20081} there are only a few, rather old, papers reporting cross section measurements of the DEA process as a function of the incident electron energy. We cite here Rapp and Briglia~\cite{rapp:1480}, Schulz~\cite{PhysRev.128.178} and Christophorou {\it et al.}~\cite{:/content/aip/journal/jcp/43/12/10.1063/1.1696685}. DEA from the $^1\Delta_g$ state of O$_2$ has also been observed~\cite{Jaffke199262} and found to have a cross section of similar magnitude to process (\ref{eq:DeA_process}). Cross sections for EID only appear to have been measured for electronically excited states of O$_2$ by Cosby~\cite{Cosby93}.

In Born-Oppenheimer approximation, the dynamics of the oxygen nuclei in DEA is treated within the local-complex-potential model~\cite{Domcke199197}, already satisfactory adopted for resonant calculations in other diatomic molecules (see Refs.~\cite{0963-0252-22-2-025001,N2_res_diss} and references therein). The corresponding cross section from the vibrational level $v$ of oxygen and for electron energy $\epsilon$ is given by:
\begin{equation}
\sigma_v(\epsilon) =
2\pi^2\,\frac{m_e}{k}\,\frac{K}{\mu}\,\lim_{R\to\infty}\left|\xi(R)\right|^2\,,
\label{eq:DeA_cs}
\end{equation}
where $K$ is the asymptotic momentum of the dissociating fragments O and O$^-$ with reduced mass $\mu$; $m_e$ and $k=\sqrt{2m_e\epsilon}$ are the incoming electron mass and momentum respectively and $\xi(R)$ is the solution of the Schrodinger-like equation for the resonant state and total energy $E=\epsilon_v+\epsilon$:
\begin{equation}
\left(-\frac{\hbar^2}{2\mu}\frac{d^2}{dR^2} + V^- + \frac i2\Gamma -
E\right)\xi(R) = -V\,\chi_v(R)\,,\label{eq:nuclearmotion}
\end{equation}
where $V^-+\frac i2\Gamma$ is the complex potential of the resonance reported in Fig.~\ref{fig:pes}, $V^2=\Gamma\,/(2\pi\, k)$ is the discrete-to-continuum potential coupling and $\chi_v$ is the vibrational wave function of O$_2$ corresponding to the vibrational level $v$. $R$ represents the internuclear distance. There is not interference among the resonant states with different symmetries and for the two $\Sigma^-_u$ states the interference can be considered negligible as they have different spin multiplicity.

\section{\label{R_D} Results and discussion}
\begin{figure}
\includegraphics[scale=.9]{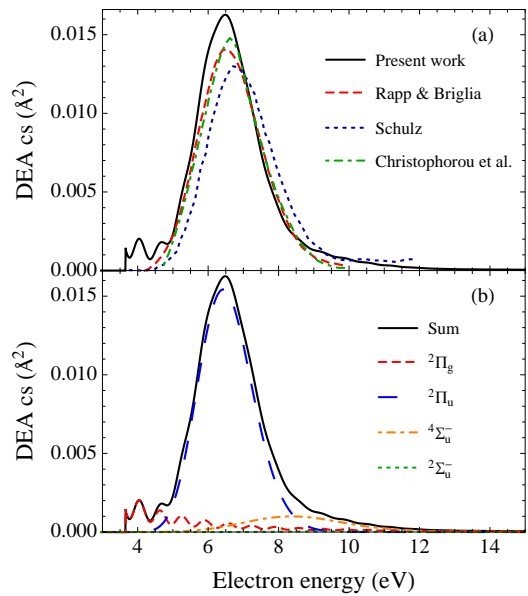}
\caption{(Color online)  (a) Calculated dissociative electron attachment  cross section
for $v=0$ compared with experimental results of Rapp and
Briglia~\cite{rapp:1480}, Schulz~\cite{PhysRev.128.178} and Christophorou
\emph{et al.}~\cite{:/content/aip/journal/jcp/43/12/10.1063/1.1696685}. (b)
Contributions from the four resonances to the total cross section.
\label{fig:dea_cs}}
\end{figure}

Figure~\ref{fig:dea_cs}(a) shows our results for the DEA cross section for O$_2$($v=0$) compared with the experimental results of Rapp and Briglia~\cite{rapp:1480}, Schulz~\cite{PhysRev.128.178} and Christophorou \emph{et al.}~\cite{:/content/aip/journal/jcp/43/12/10.1063/1.1696685}. The agreement is excellent, within the experimental error of $\pm15\%$: the peak is positioned at 6.43~eV with an absolute value of 0.0154~\AA$^2$ and the FWHM of about 2~eV.  Fig.~\ref{fig:dea_cs}(b) reports cross sections coming from all four resonances and their sum. As found experimentally, the main contribution to DEA cross section comes from the $^2\Pi_u$ state of O$_2^-$. We also note, at about 8.5~eV, the presence of a significant contribution due to the 
$^4\Sigma^-_u$ symmetry which, added to the main $^2\Pi_u$ contribution, reproduces the high-energy tail observed experimentally. The non-negligible contribution of $^4\Sigma^-_u$ resonance to DEA cross section has already been deduced from the measured angular distribution of O$^-$ ions~\cite{0953-4075-39-14-L01}.

\begin{figure}
\includegraphics[scale=.8]{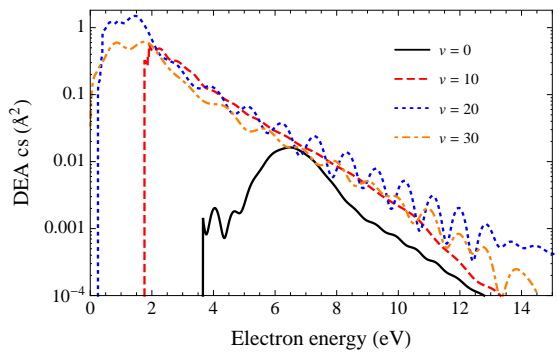}
\caption{(Color online)  Calculated dissociative electron attachment cross sections for vibrationally excited O$_2$, for levels $v=10$, 20 and 30, compared with the ground level $v=0$. The cross sections include all the four resonant contributions. \label{fig:dea_cs_v}}
\end{figure}

Figure~\ref{fig:dea_cs_v} shows our DEA cross sections calculated for higher vibrational levels of O$_2$, $v=10, 20$ and 30, compared with the result $v=0$. All cross sections are summed over all four resonance states.  
As expected~\cite{SandT_NJP}, the threshold of the process shifts to lower energies as the vibrational level increases; 
the is due to the reduced threshold for dissociation limit. 
At same time the maximum value of the cross section for $v<30$ grows by of several orders of magnitude at low energies. This behaviour is due to the survival factor $e^{-\rho}$~\cite{PhysRev.155.59}, given approximatively by:
\begin{equation}
e^{-\rho}\simeq \exp\left[-\int_{R_e}^{R_c}\frac{\Gamma(R)dR}{\hbar
v(R)}\right]\,,\label{eq:survival}
\end{equation}
where integration is extended over the region between the classical turning point ($R_e$) and the stabilization point ($R_c$) and $v(R)$ is the classical velocity of dissociation. As the vibrational level increases, $R_e$ increases and the survival probability grows like the cross section. The oscillations, visible at high vibrational levels, results from the interplay between the neutral vibrational wave function and the resonant continuum wave function. Analysis of the resonance contributions to the cross section at $v=20$ shows a complicated picture with both $\Pi$ resonances making significant contributions at low energy and important contributions from the $\Sigma$ resonances at higher energies.

Figure~\ref{fig:dea_cs_v} also shows the cross sections for $v=30$. The increasing trend of the maxima is now inverted. The eigenvalue for this vibrational level~\cite{0963-0252-22-2-025001}, in fact, lies above the energy of the $\textrm{O}^-(^2\textrm{P}) + \textrm{O}(^3\textrm{P})$ asymptotic state (see Fig.~\ref{fig:pes}), so that the DEA process becomes exothermic with a threshold at zero-energy. For this case $R_c>R_e$ so that the exponent $\rho$ in Eq.~(\ref{eq:survival}) vanishes ($\Gamma(R>R_c)=0$) and the magnitude of the cross section is no longer governed by the survival factor.  The decreasing trend of the cross section maxima is maintained for $v>30$.

The corresponding EID cross section from $\textrm{O}_2(\textrm{X}\,^3\Sigma^-_g; v)$ and electron energy $\epsilon$ is given by~\cite{N2_res_diss}:
\begin{equation}
\sigma^{\textrm{EID}}_v(\epsilon) = \frac{64\pi^5m^2}{\hbar^4}\,\int
d\epsilon'\frac{k'}{k}\left|\left\langle \chi_{\epsilon'}(R)|V|\xi(R)
\right\rangle\right|^2\,,
\end{equation}
where $\langle\cdots\rangle$ means integration over the internuclear distance $R$, $\xi(R)$ is the resonant wave function solution of the Eq.~(\ref{eq:nuclearmotion}) and $\chi_{\epsilon'}$ is the continuum wave function of O$_2$ with energy $\epsilon'$ representing the 2O + $e^-$ fragments. The continuum energy $\epsilon'$ was integrated from O$_2$ dissociation threshold up to 10 eV.

\begin{figure}
\includegraphics[scale=.8]{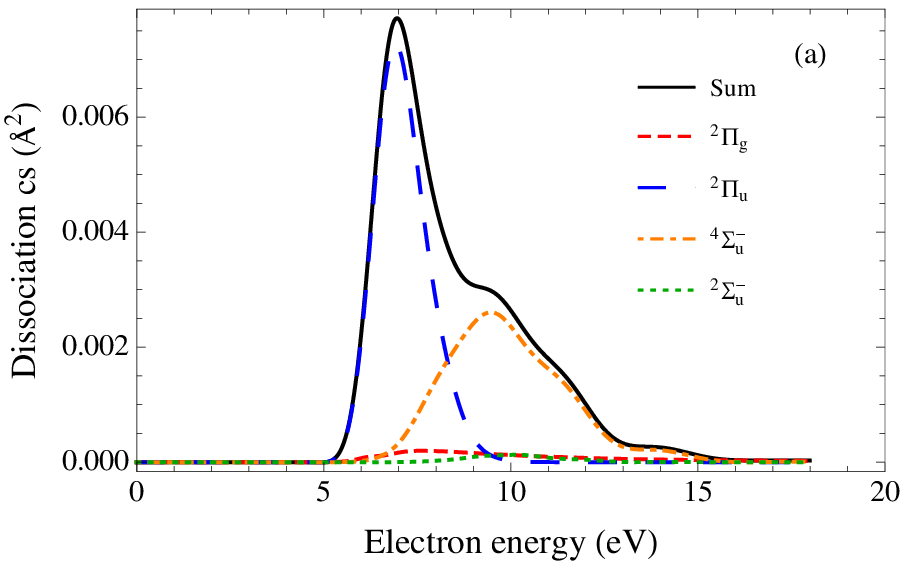}
\\
\includegraphics[scale=.8]{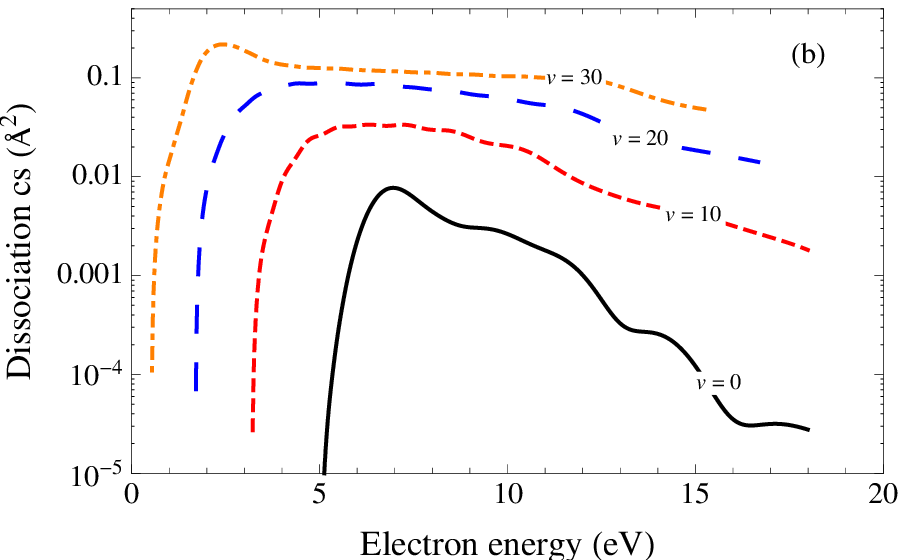}
\caption{(Color online)  (a) Calculated cross sections for resonant electron impact dissociation
of oxygen for $v=0$. Contributions from the four O$^-_2$ resonances are shown;
the black curve is the sum. (b) Dissociating cross sections for vibrationally
excited oxygen at $v=10$, 20 and 30 compared with the results
for $v=0$. \label{fig:diss_cs_v}}
\end{figure}

Figure~\ref{fig:diss_cs_v}(a) shows the calculated EID cross sections for $v=0$. Contributions coming from the four O$_2^-$ resonances are shown. Like DEA, the main contribution comes from the 
$^2\Pi_u$ symmetry, which gives a maximum at 6.93~eV with an absolute value of 0.0072~\AA$^2$, and a significant contribution from the $^4\Sigma_u^-$  at 9.47~eV with a cross section of 0.0026~\AA$^2$. The contribution from the other symmetries is negligible. Figure~\ref{fig:diss_cs_v}(b) shows the cross sections for excited vibrational levels of oxygen compared with those for the ground level. Cross sections include all four resonance contributions. As expected~\cite{SandT_NJP} the EID cross section increases rapidly with vibrational excitation and the threshold moves to lower energy. The cross section for $v = 30$, for example, reaches a value comparable with that of the well-known Shuman-Runge dissociative transitions for the same vibrational level~\cite{Laricchiuta2000526}, whose dissociative cross sections decrease with increasing the vibrational excitation of the molecule. We note that EID is known to occur efficiently \emph{via} electron impact excitation of  O$_2$ \cite{itikawa:1, McConkey20081}. However the threshold for electronic excitation processes lie significantly higher than those considered here, making them relatively unimportant in low temperature plasmas.

\section{\label{Con} Conclusions}
In conclusion, we present new calculations for dissociative electron attachment for oxygen using \emph{ab initio} potential energy curves and the first calculations for resonant electron-impact dissociation of oxygen. We confirm the dominant contribution of $^2\Pi_u$ symmetry in both processes starting from $v = 0$. Both cross sections however increase rapidly with increase in the vibrational excitation of  the molecule and for these vibrationally excited states it is necessary to consider the contributions from all four of the low-lying O$_2^-$ resonance states.

\begin{figure}[b]
\includegraphics[scale=.9]{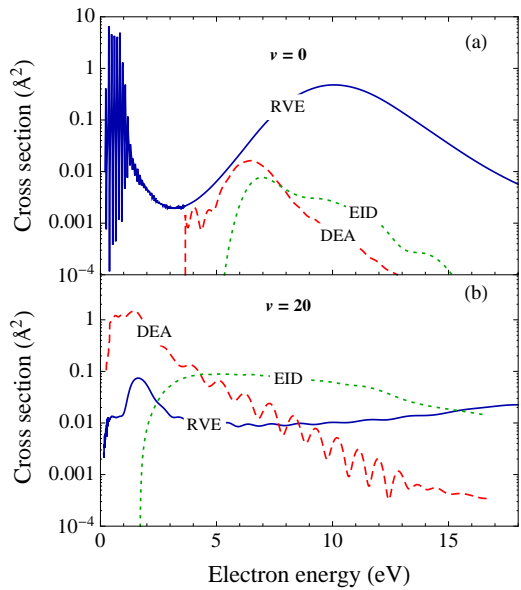}
\caption{(Color online)  Summary of the cross sections for electron-O$_2$ collisions:
Resonant vibrational excitation for $\Delta v = 1$
transition~\cite{0963-0252-22-2-025001} (solid line), dissociative electron
attachment (dashed line) and resonant dissociation by electron impact (dotted
line) for (a) $v=0$ and (b) $v=20$. The cross sections are the
sum over the four O$_2^-$ resonances. \label{fig:allxc}}
\end{figure}

Finally, Fig.~\ref{fig:allxc} compares cross sections for different resonant processes involving electron-$\textrm{O}_2(\textrm{X}\,^3\Sigma^-_g,v)$ scattering: our previous results for $v\to v+1$ vibrational-excitation~\cite{0963-0252-22-2-025001}, with the present cross sections for DEA and EID for the cases when the O$_2$ is initially in its vibrational ground state and in its $v=20$ state. For the $v=0$ level vibrational excitation is the dominant low-energy process. However for vibrationally excited molecules the DEA increases rapidly and becomes much the most important process at low energies. We  expect that the inclusion of these results will have important consequences in models of plasmas containing molecular oxygen and, in particular, will lead to a significant increase in O$^-$ ion production in these models.

Recently, we have calculated also the rate coefficients for the processes (\ref{eq:DeA_process}) and (\ref{eq:Diss_process}) as a 
function of the electronic temperature and for all the vibrational levels. Details of the calculations and results will be reported in \cite{RGD29}. DEA and EID cross sections as a function of electron energy and O$_2$ vibrational state, along with the corresponding rate coefficients, have been made freely available in the Phys4Entry database~\cite{F4Edatabase}.

\section*{Acknowledgements}
The authors are grateful to Prof. M. Capitelli (Universit\`a di Bari and IMIP-CNR Bari, Italy) for useful discussion and comments on the manuscript. This work was performed as part of the Phys4Entry project under EU FP7 grant agreement 242311. One of the authors, RC, would like to acknowledge financial support from MIUR-PRIN 2010-11, grant no. 2010ERFKXL.


\end{document}